# Engines of Power: Electricity, AI, and General-Purpose Military Transformations

Jeffrey Ding and Allan Dafoe[*]

## Abstract

Major theories of military innovation focus on relatively narrow technological developments, such as nuclear weapons or aircraft carriers. Arguably the most profound military implications of technological change, however, come from more fundamental advances arising from "general purpose technologies" (GPTs), such as the steam engine, electricity, and the computer. With few exceptions, political scientists have not theorized about GPTs. Drawing from the economics literature on GPTs, we distill several propositions on how and when GPTs affect military affairs. We call these effects "general-purpose military transformations" (GMTs). In particular, we argue that the impacts of GMTs on military effectiveness are broad, delayed, and shaped by indirect productivity spillovers. Additionally, GMTs differentially advantage those militaries that can draw from a robust industrial base in the GPT. To illustrate the explanatory value of our theory, we conduct a case study of the military consequences of electricity, the prototypical GPT. Finally, we apply our findings to artificial intelligence, which will plausibly cause a profound general-purpose military transformation.

[*] Jeffrey Ding is a PhD candidate in the Department of Politics and International Relations, University of Oxford, and a researcher at the Center for the Governance of Artificial Intelligence (GovAI). He is currently a predoctoral fellow at the Center for International Security and Cooperation at Stanford University, sponsored by Stanford's Institute for Human-Centered Artificial Intelligence. Allan Dafoe is an Associate Professor at the University of Oxford and the Director of GovAI, at the Future of Humanity Institute. Correspondence email: jding99@gmail.com.

For helpful comments and input, we thank: Ashwin Acharya, Markus Anderljung, Carolyn Ashurst, Ben Buchanan, Max Daniel, Ben Garfinkel, Charlie Giattino, Hamish Hobbs, Alex Lintz, Luke Muehlhauser, Carina Prunkl, Toby Shevlane, Waqar Zaidi, and especially Alexis Carlier and Joslyn Trager.

# I. Introduction

*AI is the new electricity.*
— Andrew Ng, Leading AI researcher and Co-Founder of Google Brain[1]

It is now standard for social scientists, policymakers, and machine learning experts to compare artificial intelligence (AI) with electricity, the quintessential general-purpose technology (GPT). International relations scholars that study the military implications of AI acknowledge this comparison, yet they have done little systematic research into the military implications of GPTs. Much work treats AI as a relatively narrow technological advance, in the mold of nuclear weapons or aircraft carriers.[2] It does not reckon with AI as a GPT, such as electricity or the computer, which is differentiated by its pervasiveness, scope for continual improvement, and strong synergies with other technologies. The comparison is gestured at but not seriously examined.

How do GPTs, like electricity and AI, influence the military balance of power? Taking military effectiveness as an analogue for economic productivity,[3] we extend insights on GPTs and economic transformations to the implications of GPTs for military transformations. Differing from narrower technologies, GPTs influence military effectiveness through a protracted, gradual process that involves a broad range of military innovations and overall industrial productivity growth. Referring to this process as a *general-purpose military transformation* (GMT), we identify three features of GMTs which relate to their breadth of impact, timeline of widespread adoption, and indirect productivity spillovers.

We then argue that GMTs differentially advantage militaries that are connected to a robust industrial base in the associated GPT, which we label "the industrial dependency hypothesis." If GPTs were like other narrower technologies, differentials in military adoption may be more tied to other factors, such as the fit between a single military innovation and a military's tactical incentives, financial resources, and organizational

---

capital.[4] These can affect the adoption of *particular* military innovations linked to a GMT, but a military's ability to draw on a robust industrial base in the GPT affects *all* military innovations linked to a GMT. To effectively exploit a GMT, militaries must draw on talent, industry, and infrastructure in the civilian realm, where the momentum for a GPT's development lies.

To empirically support our reasoning about GMTs, we examine the evolution of electricity in military affairs. Surprisingly, very little scholarship directly examines the military consequences of electricity — widely recognized as one of the most significant technological innovations in history.[5] In the 1870s and 1880s, a cluster of electrical innovations, including the electric dynamo (1871) and the steam turbine generator (1884), helped create a versatile energy system with many industrial applications in lighting, communications, transportation, and machinery. A GPT was born. Playing a notable role in World War I, military applications of electricity increased in breadth and depth throughout the interwar period. After World War II, informed observers ranked electricity among the three great influences on naval warfare in the 20th century.[6] Eliminating the "naval" qualifier would not be a stretch.

The process of military electrification exhibits the three theorized features of a GMT. First, the impact of electricity on military power materialized through a broad range of military applications, including communications (e.g. wireless telegraphy for battlefleet coordination), fortifications (e.g. searchlights to defend against night attacks), transportation (e.g. diesel-electric propulsion for submarines), and weapon systems (e.g. electric fire control). Second, electricity significantly upgraded industrial productivity, which increased military production potential. Third, similar to the slow progression of electrification across economic sectors, the spread of electrical innovations across military branches and divisions took many decades.

In line with the industrial dependency hypothesis, the extent to which militaries took full advantage of electrification depended on their connection to a robust base of electrical talent, industry, and infrastructure in the civilian economy. As established by the three features highlighted above, incorporating a GMT requires much more than acquiring a single electric dynamo or adopting one military innovation. The Russian military, for instance, pioneered the use of electronic countermeasures in combat in 1904. However, unlike leaders in military electrification like Britain, Russia's weak industrial base of electrical technology prevented it from taking full advantage of the electrification.[7] To separate the effects of the GPT dimension from other factors that could explain this military electrification gap, we assess Russia's ability to keep pace with Britain in adopting submarines, a non-GPT.

---

[4] Gilli and Gilli 2014; Horowitz 2010
[5] Exceptions include: Hezlet 1975, Headrick 1991.
[6] Sir Henry Tizard, Chairman of the UK's Defence Research Policy Committee, declared that the three great influences on naval warfare of the 20th century were the aircraft, the submarine, and the discovery of electromagnetic waves. Hezlet 1975.

This article directly engages with key academic and policy debates. First, the existing literature on military innovation overwhelmingly draws on case studies of relatively narrow technological developments, such as new weapon systems.[8] Yet, historians generally recognize that the most profound impacts of technological change arise from more fundamental breakthroughs, including GPTs like the steam engine or the computer. In addressing this gap, our study of electricity broadens the universe of cases for investigating the military implications of technological change.

Second, we develop the first comprehensive theory that sets forth the conditions under which GPTs alter the military balance of power. The limited literature on GPTs and military power posits that because GPTs are characterized by private sector dominance and relatively low fixed costs, they rapidly diffuse from technological leaders to laggards, thereby narrowing the gap in military capabilities.[9] We argue, instead, that technological laggards do not inevitably catch up in GMTs. Rather, the protracted process of GMT diffusion differentially advantages militaries which can tap into a robust industrial base in the GPT. By highlighting the unique effects of GPTs, we contribute to a growing approach to studying the impacts of emerging technologies on international security, which focuses on particular technological dimensions such as complexity,[10] disruptiveness,[11] and dual-use.[12]

---

[7] Russia's per capita production of electricity did not reach UK 1910 levels until 1930. Our calculations are based on the Cross-country Historical Adoption of Technology (CHAT) dataset by Comin and Hobjin 2009.
[8] For instance, a systematic review of 60 different cases of military innovation, sourced from 73 books and articles on the subject, does not mention electricity (Horowitz and Pindyck 2020).
[9] Drezner 2019, 300; Horowitz 2018, 39.

Finally, our research directly bears on current debates over how AI could shift the balance of military power.[13] To date, much of the discussion emphasizes the narrow effects of specific AI applications, such as autonomous weapons.[14] That AI could influence military power through its effects on industrial productivity is rarely discussed among scholars. Some scholars also suggest that AI will significantly transform military effectiveness in a relatively short timeframe, and that AI will enable rising powers to leapfrog the U.S. in military strength. As the conclusion will show, a GMT-based approach points toward different conclusions on all these fronts.

The rest of the article proceeds as follows. The next section deduces key propositions about the impact of GPTs on military affairs by adapting insights from economic and historical studies of GPTs. Our theory of GMTs is composed of three features influencing how and when GPTs affect military affairs, along with an explanation for why GMTs differentially advantage certain militaries. Leveraging primary and secondary accounts, we then illustrate the explanatory value of our theory with a historical case study: the evolution of electricity in military affairs. Finally, we discuss the limitations of our analysis and reflect on the military implications of artificial intelligence, which is plausibly a GPT as impactful as electricity.

---

[10] Gilli and Gilli 2019
[11] Mukunda 2010; Dombrowski and Gohlz 2009.
[12] Stowsky 2004. Relatedly, discussions of "Revolutions in Military Affairs" (Krepinevich 1994) and long-wave theories. (Modelski and Thompson 1996) also contend that basic innovations in the commercial domain can drive military-technical revolutions. These accounts relate to the broad outlines of our theory, but they do not specify how and when these transformations occur. We specify GPTs as a particular class of technologies that can generate these revolutions.
[13] Garfinkel and Dafoe 2019; Horowitz 2018; Kania 2017; Payne 2018
[14] Some texts analyze autonomous weapons because they present thorny legal and ethical challenges. Scharre 2018; Bode and Huelss 2018

# II. Theory: GPTs and Military Power

## FOUNDATIONS OF GPT THEORY

The vast majority of theorizing about GPTs comes from economists and economic historians who analyze GPTs as the primary drivers of long-term economic growth. One systematic review of GPTs described the concept as "one of the most successful memes in economic history in the last decade."[15] With a few exceptions, political scientists have done much less theorizing about GPTs.[16] Per a search of the JSTOR database, the economics discipline accounted for 262 of 365 total articles that referenced "general purpose technology." The combined equivalent for both the political science and international relations field was 33 articles.[17]

Economists and economic historians largely agree that a GPT is defined by three characteristics.[18] First, GPTs offer *great potential for continual improvement*. While all technologies offer some scope for improvement, a GPT "has implicit in it a major research program for improvements, adaptations, and modifications."[19] Second, GPTs are characterized by their *pervasiveness*. As a GPT evolves, it finds a wide "variety of uses" and "range of uses."[20] A wide "variety of uses" refers to the diversity of a GPT's use cases (e.g. a computer can be used for storing information, calculating statistics, entertainment services, etc.), while a wide "range of uses" captures the range of industries and individuals that use a GPT. For instance, a screw has a wide "range of use" since it is used to fasten things together across a large swathe of productivity activities in the economy, but it does not have a wide "variety of uses."[21]

---

[15] Field 2008, 2.
[16] Drezner 2019; Horowitz 2018. Some literature on "leading sectors" also references GPTs. See, for example, Thompson 1990.
[17] JSTOR search conducted April 28, 2020.
[18] The following discussion is mostly drawn from Lipsey et al. 2005 and Bresnahan and Trajtenberg 1995. Other accounts employ similar definitions, albeit with some modification: Jovanovic & Rousseau 2005; Bresnahan 2010. For a critical view of the GPT concept, see: Field 2008;
[19] Lipsey et al. 1998, 39
[20] We agree with Cantner and Vannuccini's (2012) emphasis on the establishment of GPTs as a process unfolding in time. Thus, a GPT can be seen as an emergent property constituted by interactions between technological characteristics and institutions.
[21] Lipsey et al. 1998, 39.

Third, GPTs have *strong technological complementarities*. In other words, the benefits from innovations in GPTs come from how related technologies change in response, and cannot be modeled as a mere reduction in the costs of inputs to the existing production function. For example, the overall energy efficiency gains from merely replacing a steam engine with an electric motor were minimal. Facilitated by technological improvements in electric storage and machine tools, the main productivity benefits of factory electrification only materialized after plants transitioned their entire power distribution system to one in which machines were driven individually by electric motors.[22] Tying all three characteristics together, David describes the pattern of how a GPT spreads as an "extended trajectory of incremental technical improvements, the gradual and protracted process of diffusion into widespread use, and the confluence with other streams of technological innovation, all of which are interdependent features of the dynamic process through which a general purpose engine acquires a broad domain of specific applications."[23]

Though these characteristics of a GPT apply in all contexts, the rate at which a GPT acquires a broad domain of applications varies by country. For instance, empirical analysis has shown that the U.S. has benefited more from computer technologies in terms of economic productivity than its industrial rivals.[24] One of the key factors that explains this gap is the superior performance of the U.S. higher education system in adjusting to the increased ICT skill needs. In particular, cultivating engineering professions provides an important repository in which GPT learning accumulates and then spreads through different industries. Historically, engineering disciplines, such as computer science and electrical engineering, have developed in the wake of a new GPT.[25] The demands for this type of skill systematization and standardization are especially salient for the extensive adoption of GPTs, which "depend on widening, as opposed to deepening, the knowledge base."[26]

---

[22] Previously, factories were powered by shaft and belt drive systems, which relied on a single, central steam engine. Devine 1982.
[23] David 1990, 356. Emphasis ours.
[24] See, for example, Krueger and Kumar 2004.
[25] Rosenberg 1998a, 169.
[26] Vona and Consoli 2014, 1397. The key is to develop standardized routines that capture the practical and localized knowledge embedded in rapidly evolving technological trajectories, and then widen access to less-talented individuals (Vona and Consoli 2014, 1402-1403)

TRANSLATING GPT THEORY TO MILITARY TRANSFORMATION

By adapting insights from the literature on GPTs and economic productivity, we theorize about how GPTs transform military effectiveness (a GMT). Taking military divisions and branches (as opposed to industries or firms) as the application sectors of a GPT, we translate the pattern of how a GPT spreads across a national economy to a military context.[27] We first deductively articulate three key features of GMTs. Equipped with a more complete view of GMTs, we then pinpoint why different militaries are better able to exploit GMTs.

For our translation to work, we must first establish that the same economically transformative characteristics of GPTs also apply to military transformation. GPTs possess great potential for continual improvement, become pervasive in their wide variety and range of military applications, and have strong technological complementarities with existing military technology systems. The computer is one such example. Upon entering military systems, it continually improved along many technical dimensions, provoked significant structural changes, and eventually found a wide variety of uses across many branches and units (in battle simulations, logistics handling, and weapons control, etc)..[28]

We can differentiate GPTs from other militarily-significant innovations that meet some but not all three of the GPT criteria. The category of dual-use technologies, defined as technology which has both commercial and military applications, is an important near neighbor.[29] Some dual-use technologies, such as aircraft, are characterized by continual improvement and offer strong technological complementaries (e.g. benefits to militaries that adopted strategic bombing in World War II). Aircraft propulsion systems, however, have a limited variety and range of applications which means we should not expect them to give rise to GMTs.[30] The breadth of the transformations produced by GPTs often lead analysts to describe it as an "ization," such as the electrification, computerization, or *zhinenghua* (intelligentization, in Chinese) of the military. The computerization of the military describes the process of the computer's diffusion in the military domain. There is no equivalent for aircraft engines.

---

[27] We draw on some models of military innovation that disaggregate the military into individual units (e.g. branches, divisions, etc.) that can adopt technological capabilities at different rates. Grissom 2006.

[28] O'Hanlan (2018, 20) writes, "Computer systems have been introduced...into virtually all domains of military equipment and operations." Computers have propelled drones and robotics to the frontier of warfare. The proliferation of inexpensive computing power has improved capabilities in automated aircraft, missile defense interceptors, and guided mortars (O'Hanlan 2018, 11).

[29] All GPTs are dual-use, but not all dual-use technologies are GPTs. The diversity of potential applications for many dual-use technologies is limited. Horowitz 2020a.

Of course, GMTs differ from GPT-driven economic transformations. Unlike the market dynamics of a nation's economy, intra-military competition involves different units pursuing a nominal shared mission and budget flows from one big primary source. Military effectiveness and economic productivity are analogous but not the same; measuring the output of technological innovation in the military realm is much more difficult than in the civilian economy.[31] Despite these differences, some of the ways in which GPTs interact with military organizations follow a similar trajectory as GPTs in economic systems. In particular, the first two features of GMTs, regarding their broad impact pathway and prolonged timeline of diffusion, draw directly from stylized facts in the existing GPT literature.

Like translations of all kinds, adapting the foundations of GPT theory to military affairs requires some modifications. The military is typically reliant on the civilian economy to advance the GPT's development, particularly after its initial incubation.[32] The breadth of possible GPT applications in the economic realm far exceeds those in the military realm, so the full effect of a GMT includes the impact of GPT-induced productivity surges on military production capabilities. Moreover, the trajectory of a GMT will depend on a military's connection to the evolution of a GPT in the civilian economy. The rest of this section describes the three key characteristics of GMTs and the GMT leveling hypothesis in more detail.

---

[30] The general-purpose nature of a technology, like the level of disruptiveness or complexity associated with a technology, falls on a spectrum where the delineation between a GPT and a non-GPT is somewhat fuzzy. This discussion of aircraft, classified by some scholars as a GPT (e.g. Ruttan 2006), demonstrates that one can usefully exclude "near-GPTs" from consideration based on established criteria of a GPT. In a systematic survey of lists of historical GPTs, none of the texts included aircraft as a GPT. (Field 2008)

[31] Rosen 1994, 46.

[32] Some scholars have shown that military investment played a crucial role in incubating some GPTs. See Misa 1985; Ruttan 2006; Smith 1985. Misa's work tracks how the U.S. Army Signal Corps served as an important institutional entrepreneur in the development phase of the transistor. Military investment, however, is not *necessary* for seeding GPTs, as the history of commercially-initiated developments in steam engines and electricity show (Bresnahan and Trajtenberg 1995, 95-96).

GMTS: THREE FEATURES OF GPT TRAJECTORIES IN MILITARY AFFAIRS

**First, GPTs directly enhance military power by spurring a wide range of military innovations.**

When scholars analyze the impact of technology on warfare, their attention often goes to the most visible and graphic part of warfare: the projectile or other mechanism of force.[33] The impact of a weapon system is thought to materialize through a relatively narrow impact pathway. In contrast to weapons technology, which do not have many other downstream applications, GPTs influence military effectiveness through a very broad pathway. Since GPTs are utilized in a variety of ways across the entire military, they produce many downstream applications. Some may appear insignificant (the use of a computer to calculate figures for a military unit's budget), while others seem more revolutionary (the use of a computer to crack the encryption of enemy communications). The impact of a GPT on military power, therefore, depends on the distribution over all these downstream applications.

Given the broad impact pathway of a GPT, in the early stages of a GPT's development the foreseeability of a GPT's effect on the conduct of warfare is very limited. GPTs will affect military affairs in unanticipated ways. Because the advance of a GPT depends on many complementary innovations and adaptations, its full impact will not be apparent until one fully accounts for the many, lengthy causal chains shaping various sectors, and the ways those sectors adapt. For example, when the steam engine was invented in the eighteenth century, it was initially used exclusively to pump water out of flooded mines. It was not until the early nineteenth century that the steam engine became a generalizable source of power for factories, railroads, and naval ships.[34]

A certain degree of unpredictability applies to the effects of all innovations on military affairs. But the breadth of military applications affected by an innovation can vary. Some innovations present a very limited set of applications, though these applications can interact with the strategic landscape in many ways

---

[33] Beckley 2010, 55.
[34] Rosenberg 1996, 345. The steam engine is widely recognized as a GPT. (Field 2008); Predicted demand for computers was also very off-base. A survey by the Dept. of Commerce conducted in the late 1940s estimated that about 100 mainframe computers would satisfy the entire needs of the nation. Helpman and Trajtenberg 1996, 30 (fn 16)

(Figure 1). For instance, nuclear fission technology had a relatively bounded set of military applications — namely, nuclear weapons — but the interaction between nuclear weapons and the strategic landscape evolved in multifaceted, unpredictable ways.[35] Our argument is that the set of possible military applications for GPT innovations is much larger than the corresponding set for other technologies, thereby severely limiting the foreseeability of its military implications.

*Figure 1: The Military Technology Stack — Two Impact Pathways*

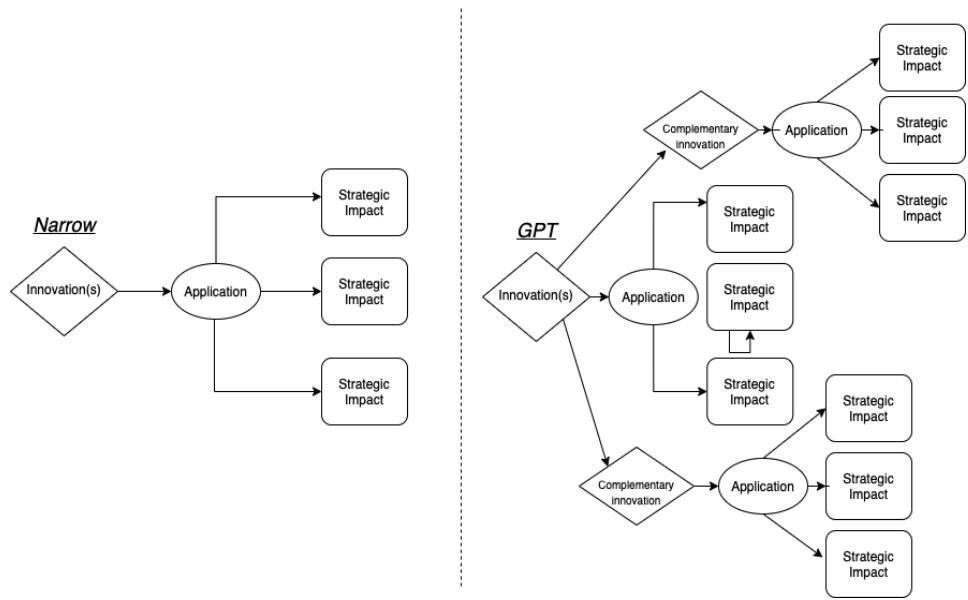

***Second, GPTs indirectly affect military power by significantly upgrading industrial production potential.*** In addition to the direct impact pathway, it is necessary to consider the ramifications of GPTs for a country's production potential. Many scholars have noted that the productive capacity of a nation's industrial base undergirds military power.[36] Across hundreds of battles and wars from 1898 to 1987, Beckley has found a positive relationship between economic productivity and military effectiveness. Separate from its total economic output, a more economically efficient nation will translate its economic resources into superior weapons and military organization.[37]

---

[35] Morgenthau 1964
[36] Kennedy 1987, Gilpin 1975; Kirshner 1998. Some previous scholarship on RMAs does discuss the indirect effects of technological change on "production revolutions" — e.g. the effects of the American System of Manufactures on the World War I RMA. But this is not a regular feature of RMAs. Most RMAs, including the prototypical example of the nuclear revolution, do not involve a substantial upgrading of the industrial base. This marks another clear distinction between GMTs and RMAs.

Again, this feature of GMTs derives from the fact that GPTs differ from other technologies. Assessed on their own merits alone, even the most transformative (non-GPT) technological changes do not tip the scale far enough to significantly affect overall industrial productivity.[38] As "engines of growth," GPTs are different because their impact on productivity comes from accumulated improvements across a wide range of complementary sectors; in other words, they cannot be judged on their own merits alone.[39] According to empirical studies of the steam engine, electricity, and information and communications technologies (ICTs), the episodic arrival of GPTs precedes a wave of economy-wide productivity growth.[40]

*Third, the most consequential impacts of a GPT on military effectiveness occur only after a long period of gestation.* Many decades pass between the emergence of a GPT and its substantial effect on productivity. In the 1980s economists who had pointed to GPTs to explain changes in long-term economic growth were confronted with a paradox. Transformative GPTs like computers were claimed to greatly increase productivity, yet "we (saw) the computers everywhere but in the productivity statistics."[41] Evidence of a long diffusion lag — "a time to sow" and "a time to reap" in the formulation of Helpman and Trajtenberg — between the emergence of a GPT and its full effects on productivity helped address this paradox.[42]

Compared to other technologies, GPTs exhibit a much more pronounced diffusion lag due to their substantial demands for complementary innovations, organizational changes, and upgrading of technical skills.[43] Within the first decade of the introduction of the transistor, for instance, computers and hearing aids had already incorporated semiconductors. But the telecommunications sector took much longer to adopt this GPT, despite the fact that the development of semiconductors was largely inspired by the potential benefits for the telecommunications sector.[44] Though the first dynamo for industrial application was introduced in the 1870s, the full impact of electricity on overall manufacturing productivity did not occur until the 1920s. This was only realized after organizational adaptations, such as changes in factory layout, and complementary innovations, such as the steam turbine, which enabled central power generation in the form of utilities.[45]

---

[37] Beckley 2010; See also Beckley 2018.
[38] Fogel's classic study of the social savings linked to railroad construction in the U.S. led him to conclude that "the railroad did not make an overwhelming contribution to the production potential of the economy." Fogel 1964, 235.
[39] Bresnahan and Trajtenberg 1995.
[40] Ruttan 2006, 5; David 1990; Brynjolfsson et al. 2017
[41] Robert Solow's famous quip in 1987
[42] Helpman and Trajtenberg 1994
[43] Brynjolfsson et al. 2017; David 1990

The nature of GPTs raises difficult questions about when their most consequential military effects materialize. For example, one could trace present-day advances in military electronics and precision warfare back to 19th-century electrical innovations. When does a GMT end? For our purposes, we date this to when a GPT spreads across a wide range of military applications.[46] Oftentimes, another GPT, such as the emergence of the transistor in the 1940s, opens up a new round of general purpose military transformation. Thus, advances in military electronics and precision warfare, though built on a base of electrical advances, are more connected with the transistor GMT than the electrical one.

## GMTS AND DIFFERENTIAL ADVANTAGES

We hypothesize that militaries more connected to a strong industrial base in the GPT are better positioned to exploit GMTs. The broad applicability of GPTs across many sectors, combined with the fact that the civilian economy presents many more application scenarios than the military realm, means that the momentum for a GPT's development lies in the civilian realm.[47] This distinguishes GPTs from (non-GPT) dual-use technologies, such as nuclear power.[48] To bring about a GMT, therefore, a military will need to rely more on the civilian sector for flows of knowledge, talent, investment, and complementary innovations. This leads to the following industrial dependency hypothesis: *A GPT differentially advantages militaries that can draw from a robust industrial base in the GPT.*

---

[44] Recall that the primary task of Bell Labs, the birthplace of semiconductors, was to develop the telecommunications systems manufactured by AT&T. Bresnahan and Trajtenberg 1995.
[45] Smil 2005, 33-97. David 1990.
[46] In the electricity case, this occurs during World War II.
[47] Earlier in this section, we acknowledge that in some cases the military has played an important role in initiating developments in GPTs.
[48] Maintaining a strong nuclear weapons capability does not depend on the entire industrial base's strength in civilian nuclear applications.

This proposition differs from arguments that GPTs level the military balance of power. Drezner compares GPTs, which he characterizes by private sector dominance and low fixed costs, to prestige tech like space exploration programs, defined by public sector dominance and high fixed costs. Arguing that GPTs rapidly diffuse from technological leaders to laggards, he posits that "general purpose tech has a greater leveling effect than prestige tech."[49] Horowitz also links commercially-driven GPTs to a leveling effect. Specifically, he describes this effect for AI:

> "If commercially-driven AI continues to fuel innovation, and the types of algorithms militaries might one day use are closely related to civilian applications, advances in AI are likely to diffuse more rapidly to militaries around the world...The potential for diffusion would make it more difficult to maintain 'first-mover advantages' in applications of narrow AI. This could change the balance of power, narrowing the gap in military capabilities not only between the United States and China but between others as well."[50]

Evaluating these hypotheses will depend on how one conceptualizes the acquisition of a GPT. In our view, a military's acquisition of a single electric dynamo should not count as successful diffusion, as this does not capture whether the military has meaningfully incorporated the GMT associated with electricity. That GPTs are driven forward by civilian applications does not necessitate that they will diffuse quickly to militaries that are technological laggards. Instead, the protracted, challenging process of a GMT differentially advantages militaries able to tap into a robust industrial base in the associated GPT.

The emphasis on an industrial base with specific competencies in the GPT also distinguishes this hypothesis from existing work on the economic roots of military effectiveness. Scholars have shown that the level of economic development has a significant effect on military effectiveness, reasoning that states with more efficient economies can field skilled personnel that can absorb complicated technology effectively.[51] We propose that, even among developed states, the connection to the industrial base's development of a GPT has a significant effect on which militaries take better advantage of GMTs.

---

[49] Drezner 2019, 300.
[50] Horowitz 2018, 39. To be clear, this is only one such scenario Horowitz outlines. Countervailing factors in favor of non-leveling in his careful assessment include: the possibility that the most important specific military uses of AI are exclusively based in military research and the constraint of compute costs, which could price out all but the wealthiest countries from adopting higher-end AI capabilities.
[51] Beckley 2010

# III. Case Study: The Electrification of Warfare (late 19th and early 20th centuries)

RESEARCH DESIGN AND CASE SELECTION STRATEGY

To evaluate our theory, we conduct a historical case study of the impact of electricity on military affairs. There is some debate over which technologies qualify as GPTs, but electricity is unanimously viewed as a prototypical GPT.[52] This makes it a typical, or representative, case for studying GMTs. The frequent contemporary comparisons made between AI and electricity serve as an additional advantage. Studying the electrification of warfare is also substantively important in its own right. Despite its wide recognition as one of the most significant technological changes in history, the electrification of the military has not received much scholarly attention.[53]

If the evidence from the electricity case supports our theory, we should observe two main sets of implications. First, the impact of electricity on military effectiveness should exhibit the three features of a GMT: *broad impact pathway, indirect productivity benefits, and prolonged gestation period*. Second, in line with the *industrial dependency hypothesis*, we should see differentials in military electrification based on militaries' connections to a robust industrial base in the GPT (*industrial dependency hypothesis*).

To control for competing explanations of how GPTs affect the military balance of power, we compare military electrification to the impact of the submarine on military affairs. These two technologies differ with respect to whether they qualify as a GPT but are similar across most other relevant features.[54] The submarine was not a GPT. While it generated some technological complementarities with advances in weapon systems and underwater propulsion, submarine technology did not have a wide variety and range of uses and had limited potential for continual innovation. As Michael Horowitz writes, "[AI] is an enabler, a general-purpose technology with a multitude of applications. That makes AI different from, and broader than, a missile, *a submarine*, or a tank."[55]

---

[52] Ristuccia and Solomou 2014, 227. Historians disagree about which technologies qualify as GPTs, but almost all lists include the steam engine, electricity, and information and communications technology (ICTs). Field 2008, 10.
[53] Exceptions are Hezlet 1975 and Headrick 1991, which we draw from in our empirics. Headrick limits his analysis to electrical communications, and Hezlet only covers the effect of electricity on naval warfare.
[54] For more on most-similar-system comparisons, see Beach and Pederson 2016, 239-240.

      The similarities between the submarine and electricity cases across other features help test GMT theory against alternative explanations. One might posit that the divergent outcomes of the cases, as they pertain to GMTs, are due to differences in the disruptiveness of the two technologies. Yet the submarine was at least as disruptive as electricity. The submarine is often described as a major military innovation that changed the conduct of warfare.[56] Moreover, the introduction of submarines into military affairs occurred in the same period as the introduction of electricity into military affairs.[57] If we find that the two technologies affected the military balance of power in different ways, then these differences cannot be accounted for by technology-agnostic factors, such as the organizational competencies and cultures of the militaries in question, and time-dependent factors, such as the distribution of military power and the nature of military competition at the time. These overlapping trajectories help pinpoint the technological dimension of GPTs.

## THE ELECTRIFICATION OF WARFARE — BACKGROUND

      Much of our empirical analysis comes from primary sources, supplemented by secondary accounts of the history of electricity development and military modernization. From the 1870s, the establishment of trade journals such as *The Electrician* (London) and *The Electrical Review* (Chicago) provided forums for electrical engineers to deliberate over the issues of the day, including electrical applications in military affairs.[58] The accounts of technical innovators involved with incorporating electricity into military operations are also valuable sources. One important figure was B.A. Fiske, who served as Rear Admiral of the U.S. Navy and invented more than 130 electrical and mechanical devices.[59] Records of Fiske's lectures and writings during the period, in addition to his autobiography, provide a firsthand evaluation of the uses of electricity in war.

---

[55] Horowitz 2018. Emphasis ours.
[56] Lautenschlager 1986, 121.
[57] In fact, before WWI, advances in the capacity of electric storage batteries expanded the submergence period of submarines, making submarines an application sector of electricity as a GPT. The first electrically propelled submarines appeared in the mid-1880s. Skjong et al. 2015, 6.
[58] From 1872 to 1882, 12 journals devoted to electricity sprung up in London, Paris, St. Petersburg, Berlin, New York, Chicago, and Barcelona. Manubens 2011, 171.
[59] Baines et al. 1942, 10–11. In 1882, the 28-year-old Fiske foresaw the critical need for electricity in the navy of the future. Since the navy offered no training in the subject at the time, he got permission from the Navy to study at the GE

Our analysis of the development of electricity mainly covers the period from the mid-1800s to the end of World War I. In the late 19th century a cluster of electrical inventions — including the first practicable dynamos, the transformer, and the steam turbine — enabled the widespread application of electric power. This versatile energy system transformed systems of lighting, manufacturing, transportation and electronic devices.[60] "No other technical innovation had such a far-reaching impact on modern civilization as the creation of an admirably reliable system of electricity generation, transmission, and conversion," writes Smil.[61] By tracing the evolution of military electrification across a variety of military applications, we show that the well-known versatility of electrical applications in the economic realm extended to the military domain (Figure 2).[62] This process will be further fleshed out in evaluating the four propositions.

Figure 2: The Military Electrification GPT Tree (early applications)

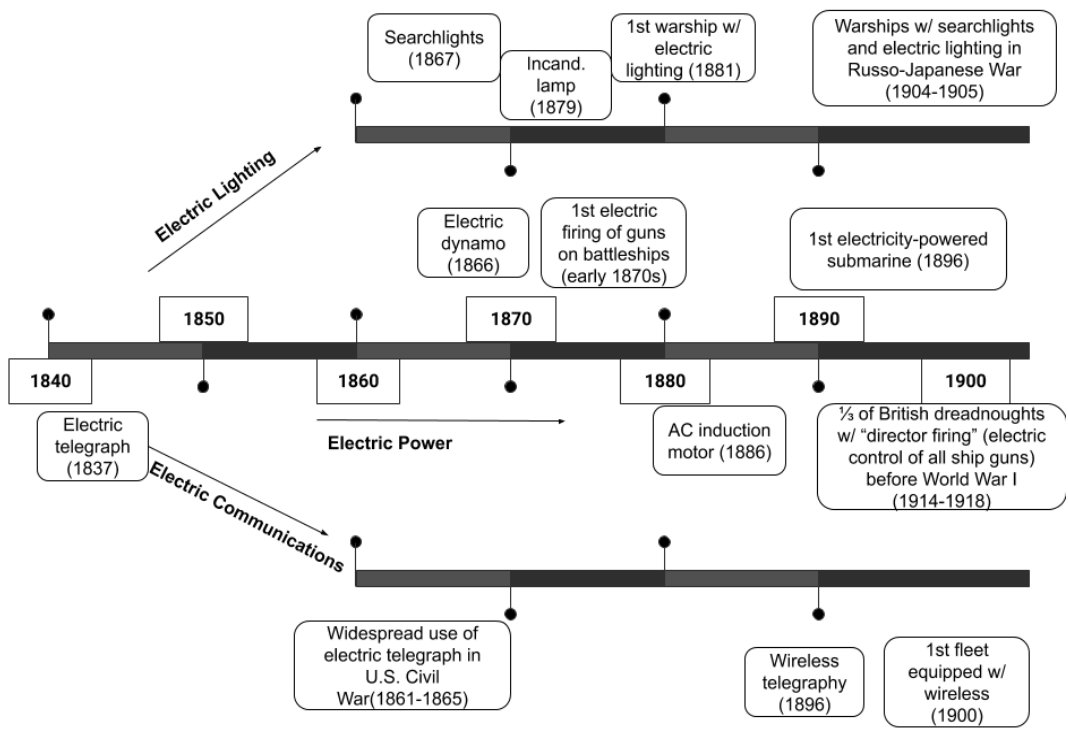

---

plant for a year. Capt. Wayne P. Hughes Jr. writes, "The Navy never got a greater payback from graduate education than that investment." Hughes Jr. 2017.

[60] Smil 2005

[61] Smil 2005

[62] Extensive citations for the dates in this figure are available in an online appendix.

GMT FEATURE 1: BROAD IMPACT PATHWAY

The main applications of electricity in the late nineteenth century were in lighting (searchlights), electric power (electric handling of weapon systems), and communications (the telephone and telegraph). Searchlights, which cast a concentrated beam of light, helped to guard harbors and forts against surprise attacks at night. First employed by the French to reveal the Prussian army's siege of Paris in the Franco-Prussian War, searchlights provided mainly defensive advantages.[63] After this episode, other nations quickly adopted searchlights into their military and naval services.[64] All the forts in New York harbor were upgraded with dynamos for searchlights in 1898.[65] Relatedly, incandescent electric light proved useful in signaling at nights, as military units could light up certain combinations of numbered lamps to warn of an impending attack.[66]

The electric firing of guns was another key component of military transformation.[67] Electric guns could be fired more quickly, accurately, and systematically. By systematic, we refer to the electric firing of guns all at once with a single trigger controlled by a central director; this trend intersected with the centralization of control in "transmitting stations" on ships which would calculate the range of projectiles with electric instruments. By 1914, around half the British battlefleet had director firing, and all had centralized control with transmitting stations.[68]

Advances in electricity fed into telegraph and telephone communications, which enabled a military system that could coordinate mass armies. "In armies, aside from submarine and subterranean mining, the principal use of electricity is in the military telegraph service, and its importance there can hardly be overestimated," commented Fiske in 1890.[69] In part due to the efficiency of its telegraph service, the Prussian Army was able to rapidly mobilize and achieve victory in 1870. The U.S. military also adopted electrical communications systems. New wireless technology improved military systems of telegraph communications, and the U.S. government purchased 1300 telephone sets for military use in 1905.[70]

---

[63] Pritchard 1877, 282; Hezlet 1975, 21. However, in some cases searchlights did facilitate offensive operations at night.
[64] Fiske 1886, 86.
[65] Marvin 1990, 145.
[66] Fiske 1886.
[67] Hezlet (1975, 25) categorizes electric firing for guns, electric lighting, and electric power for laying and training guns and supplying ammunition as "minor uses for electricity in warships" for which electricity provided minor improvements over other alternatives. The impact of GMTs comes from the accumulation of all these improvements across a wide distribution of applications.
[68] Hezlet 1975, 82.

Eventually, the electromagnetic spectrum became a new arena for conflict. Described by Rosen as "one of the most significant military technological innovations of the modern era," electronic warfare was first demonstrated in the Russo-Japanese conflict of 1904.[71] In the First World War, the French and British jammed and spoofed the electromagnetic signals of German bombing operations. Working for the British War Office, the Western Electric Company developed a method in 1915 to jam reception in German listening posts.[72] By World War I, it was difficult for ships to function without wireless communication, and this established radio intelligence as a crucial aspect of battlefield effectiveness. British radio intelligence broke German ciphers and could track German communications at sea. Thus, "radio intelligence has been assessed as the most important single factor in the defeat of the U-boats in 1914-1918."[73] Arguably the most powerful demonstration of how electricity shaped World War I was the British penetration of German diplomatic codes, which facilitated the leakage of the Zimmerman telegram.[74]

In many areas, electrical applications in communications, lighting, transportation, and weapons management converged in complete electrical systems. Consider Fiske's description of a state-of-the-art naval ship and fort in the late 19th century: "It is probable that in the near future every man-of-war and every fort will be fitted with a complete 'electrical system' well-protected from projectiles, which will include dynamos capable of supplying a very large amount of electrical energy to a system of mains, from which all the incandescent lights, all the search-lights and all the motors of different sorts can draw the supply of energy requisite for their needs."[75] To further supplement the system, electric technology controlled the firing of guns; electric-powered velocimeters would determine the exact position of an approaching vehicle; and telegraphs would communicate the commander's orders to all the departments of a ship. For *Electrical World*, the mayor of Kansas City captured the accumulated impact of electricity across a broad range of systems by 1890: "[Electricity] now not only guards the vessel from the inventions of the enemy, but aims and fires the guns, illuminates the sights that the aim may be sure, discharges torpedoes, measures her speed, is the most successful motor for submarine boats, and renders possible a system of visible telegraphy by which communications may be flashed against the clouds and understood at a distance of sixty miles."[76]

---

[69] Fiske 1890, 186.
[70] Marvin 1990, 145.
[71] The Russian military jammed the radio transmissions on Japanese battleships. Price 1984, 3-6; Rosen 1994, 190.
[72] National Electrical Manufacturers Association 1946, 87.
[73] Hezlet 1975, 143
[74] This caused the U.S. to enter the war. According to Headrick (1991, 170), "Never before or since in history has communications power been so concentrated and so effective."
[75] Fiske 1886, p. 90.

To underscore this point, electricity was at the core of the development of the dreadnought, a new type of battleship often referenced as the key military innovation of this time period.[77] Electrical communications and range finders supported *centralized fire control*, a system that facilitated effective shooting with heavy gunnery from long ranges — the crucial advance from pre-dreadnought ships to dreadnoughts.[78] Wireless telegraphy was also essential for the coordination between dreadnoughts and the overall battlefleet.[79]

Another observable implication of this broad impact pathway is that the foreseeability of electricity's impact on the conduct of warfare should have been very limited. As advances in electricity were emerging, many predicted that the main application in the military realm would come in the form of war-winning weapons. Experts and popular commentators alike envisioned military scenarios in which "some electrically ingenious device secured a strategic battlefield advantage."[80] Engines would deliver electric shocks "of infinite variety" on the battlefield, one publication hypothesized in 1889;[81] electric rays of destruction would work "revolutionary effects on the art of modern warfare," predicted another in 1896;[82] Tesla, the scientist who designed the alternating-current electric system, proposed that electric technology would facilitate remote-controlled conflict facilitated by mechanized weapons.[83] The inventor of the Gatling gun, R.F. Gatling,

---

envisioned the invention of a powerful electrical machine that could kill whole armies at the flip of a switch would bring peace to the world.[84] A 1911 edition of *Technical World* magazine painted a particularly vivid vision of what electric-powered warfare would look like in 1950:

> "The old War God hurling his thunderbolts will seem impotent beside man wielding the forces of nature for weapons. Magazines exploded without warning by darting, invisible, all-penetrating currents of electricity; devastating waves of electricity, or of some more powerful force, flashing over hundreds of miles consuming all that comes within their scourging blast. Guns, explosives, and projectiles will sink into the past, even as have the bow and arrow, giving place to howling elements clashing under man's direction."[85]

The electric rays of destruction never materialized. As Marvin summarizes, "Actual as opposed to fantasy developments in electrical warfare were mostly in the realm of communications rather than destructive weaponry."[86] Other misses were less spectacular though still significant. Even Fiske, one of the most prescient observers who played a leading role in U.S. efforts to electrify its military, severely underestimated the impact of wireless communications.[87] For example, Fiske asserted in a 1904 article that the radio had "no military usefulness whatever."[88] Fiske also expected that all-electric motors would dominate marine propulsion.[89] In actuality, mechanical drive became the dominant method of naval propulsion, though diesel-electric systems were widely adopted by submarines.[90] In an 1892 article, one American ordnance engineer, a frontline user of electrical applications, captured the limited impact predictability of electricity, "Great as was the usefulness of electricity during the period of the Civil War…it was only a means of communication; and there was probably *no one at that era whose imagination was sufficiently elastic* to dream of electricity ever acquiring the compass it possesses at the present time."[91]

---

[84] *Western Electrician (Chicago)* 1891, p. 221. These mistaken predictions of electricity in the form of a war-winning weapon still lingered in the1930s. Worried about rumors that Germany possessed a "death ray," the UK Air Ministry asked Scottish physicist Robert Watson-Watt to investigate this destructive application of electromagnetic radiation. Watson-Watt dismissed the likelihood of a death ray and responded in a memo that he was focused instead on "'the difficult, but less unpromising, problem of radio-detection as opposed to radio-destruction." APS News 2004.
[85] La Baueme 1911, 439; quoted in Marvin 1990, 144.
[86] Marvin 1990, 144.
[87] "The development of wireless was the outstanding influence of the electron on sea power in the first fourteen years of the twentieth century," writes Hezlet (1975, p. 78) One of the key figures in the development of wireless telegraphy is Guglielmo Marconi, an electrical engineer, who won the 1909 Nobel Prize in Physics for his work on the detection of electromagnetic radiation.
[88] Howeth 1963, 65; cited in Douglas 1985, 119.
[89] Fiske 1890.
[90] O'Rourke 2000.

## GMT FEATURE 2: INDUSTRIAL PRODUCTIVITY SPILLOVERS

In addition to spurring a variety of military innovations, a GPT should also transform military effectiveness by boosting industrial productivity. In the case of electricity, it is well-documented that the diffusion of electricity across manufacturing industries resulted in a productivity surge.[92] The U.S. case is extensively studied.[93] The use of electric motors expanded from slightly less than 5 percent of installed horsepower in US manufacturing in 1899 to 55 percent by 1919.[94] Crucially, electrification enabled mass production, as the adoption of electric unit drive in factories resulted in standardized workflows and plant capacity expansion.[95]

The impact of electrically-boosted production capabilities was revealed in the two great wars of the 20th century. Zeitlin concludes that, by WWI, the "capacity of civilian firms to manufacture large numbers of standardized weapons became increasingly central to the conduct of industrialized warfare."[96] The resultant increase in warmaking capacity was stark. For example, Britain possessed only 154 airplanes at the outbreak of WWI, but British aircraft factories were producing 30,000 planes per year by the end of the war.[97] Access to cheap, plentiful electricity drove these surges. From one survey of the U.S. electrical industry's contributions to WWI production, many leading electrical manufacturers estimated that 90-95 percent of their plants were supporting government needs.[98] Recognizing that the U.S. needed to realize "the greatest possible production of needed war materials of the kind peculiarly dependent upon a cheap and dependable supply of electricity," the War Industries Board placed restrictions on civilian uses of electricity in key industrial centers, so as to maximize access for war industries.[99]

---

[91] Parkhurst 1892, 359. Emphasis ours.
[92] Based on an analysis of the usage of electricity in the manufacturing industries of Britain, France, Germany, Japan, and the U.S. Ristuccia and Solomou (2014) question the productivity benefits associated with electrification. Recent research, which takes a more active measure of adoption (patenting activity vs. mere usage of electricity), suggests that there was a generalized productivity boost. See Petralia 2020 "Following the Trail of a General Purpose Technology."
[93] David 1990; Devine 1982. But a similar effect occurred in other countries. German manufacturing adopted electric power systems at a similar rate to U.S. manufacturing. From 1907 to 1933 in Germany, there was an increase of 20 to 76 percent in horsepower generated from electric power as a percentage of total horsepower in the manufacturing sector. There was still a productivity gap between the U.S. and Germany because Germany inefficiently assimilated electrical systems. Timmer et al. 2016, 880-881.
[94] Devine 1983.
[95] Rosenberg 1998b, *The Energy Journal*, 13-14.
[96] Zeitlin 1995, 47; McNeil 1982, pp. 330-331
[97] Smil 2004, 367; see also: McNeil 1982, 330. Smil writes, "In August 1914, Britain had only 154 airplanes, but just 4 years later, the country's aircraft factories were sending out 30,000 planes per year. Similarly, when the United States declared war on Germany in April 1917, it had fewer than 300 second-rate planes, none of which could carry machine guns or bombs on a combat mission, but 3 months later Congress approved what was at that time an unprecedented appropriation ($640 million or B$8 billion in 2000 dollars) to build 22,500 Liberty engines for new fighters."

This connection between the electrification of manufacturing and military production only intensified in WWII. Electricity was one of the highest targeting priorities for Allied strategic bombing efforts because of its impact upon a wide range of German industrial activities.[100] Historians attribute the outcome of the war to a large extent on the Allied capabilities in mass production.[101] Zeitlin writes, "World War II...marks the apogee of this symbiosis between mass production and military prowess."[102] In Britain, electrical manufacturers managed state-financed "shadow factories" that helped to expand aircraft production using high-volume techniques.[103] American industry produced more than 250,000 planes during World War II, which exceeded the output of Britain and Germany combined.[104] Smil concludes, "There is no doubt that the rapid mobilization of America's economic might, which was energized by a 46% increase in the total use of fuels and primary electricity between 1939 and 1944, was instrumental in winning the war against Japan and Germany."[105]

## GMT FEATURE 3: DELAYED EFFECTS

In the economic realm, scholars hold up the diffusion of electricity as an example of the long time lag between the emergence of a GPT and its significant bearings on national productivity. Measured by percentage of total installed horsepower in manufacturing industries, adoption of unit drive, and estimates of electricity's contributions to GDP growth, American electrification did not take off until the 1920s.[106] This was a full four decades after major advances like the dynamo and incandescent light bulb emerged.

---

[98] *Electrical Review* 1919, 363-364. This was a survey of American manufacturers. For similar developments in Britain, see Bangs 1917.
[99] Keller 1921.
[100] As the strategic bombing campaigns progressed, electricity fell in priority — a development later critiqued by a retrospective United States Strategic Bombing Survey: "The German electric supply system ...was extremely vulnerable to bombing attack, and, had it been attacked systematically, it would have severely crippled Germany's industrial war machine." Quoted in Kuehl 2008, 239.
[101] McNeill 1982, 355, 358-59. Kennedy 1987, 244, 248-249.
[102] Zeitlin 1995, 47. One of Germany's strategic blunders was its underestimation of "the massive potential of the United States for industrial mobilization and production." (Millet et al. 1986, 47).
[103] Zeitlin 1995, 50.
[104] Smil 2004, 367.
[105] Smil 2004, 368.

Do we find a similarly delayed timeline for the electrification of the military? Though it is more difficult to track the effects of electricity on military productivity, we can trace how and when complementary innovations in different military branches were first introduced as military capabilities, first used in warfare, and were fully adopted as a standard military capability (Table 1).[107] Combined with the fuller technology tree introduced above, this mapping exercise of two electrical military innovations shows that even the early movers did not achieve widespread adoption until right before WWI. Among later adopters, widespread adoption did not take place until the interwar period or after WWII.

| Table 1: Delayed Impact of Electrical Military Innovations | | | | |
|---|---|---|---|---|
| | Wireless telegraphy (radio) | | Electric firing of guns | |
| **Complementary innovation** | Hertz's demonstration of radio waves (1888) | | AC induction motor (1886) | |
| **Application Sector** | Navy (early) | Air Force (late) | Navy (early) | Army (late) |
| **First Introduction as Military Capability** | British fleet equipped with wireless telegraph (1900) | Planes equipped with radio at end of WWI (1916-1918) | Electric firing introduced in British navy (early 1870s) | GE develops electric-powered miniguns (1950s) |
| **First Application in War** | Russo-Japanese War (1904-1905) | WWI (but very ineffective) | World War I (1914-1918) | Vietnam War (1960s) |
| **Widespread Adoption** | British Royal Navy has "patchy global network" that supported radio communication (1914) | Germany equips air force with complete set of radio equipment (1938) | Half of British battlefleet had director firing (by 1914) | U.S. procured 10,000 miniguns during Vietnam War (1960s) |

Demand indicators, for both electrical engineering talent and electrical voltage on ships, point to a significant acceleration of military electrification around World War II. Military applications of electricity across vehicles and firearms grew so widespread before World War II that Britain established the Royal

---

[106] For share of horsepower estimates see Rosenberg 1979, 48; Devine 1982, 46-47; or adoption of unit drive, see Devine 1983; for estimates of electrification's contributions to GDP growth, see Crafts 2002.
[107] Extensive citations for the dates in this table are available in an online appendix.

Electrical and Mechanical Engineers (REME) in 1942, which was devoted to the maintenance and repair of electrical equipment. By May 1945, the unit had 158,000 officers and men.[108] In his 1960 presidential inauguration speech to Britain's Institution of Electrical Engineers, Sir Hamish D. Maclaren reviewed the trendline in electrical demand on first-class ships. In the 1880s, these were equipped with only 3 dynamos, each supplying 200 amps at 80 volts in the 1880s. After World War II, the transition from 500kW d.c. machines to 1000kW a.c. generators enabled first-class ships to meet greater electrical demand.[109] By 1960 electrical installations accounted for 30 percent of the total cost of British warships.[110]

Indeed, some of the most significant military applications of electricity did not emerge until the 1940s. The radar is the best example. Developed based on principles first seeded by Hertz's 1888 discovery of the reflective properties of electromagnetic waves, radar systems came to play a pivotal role in World War II. The U.S. Office of Scientific Research and Development, established in 1941 to coordinate scientific research for military purposes, spent more of its $457 million budget on radar and radar countermeasures ($128 million) than any other category.[111] This investment paid off. According to one estimate, Allied electronic countermeasures reduced the effectiveness of German antiaircraft by 70 to 75 percent.[112]

The delayed diffusion of military electrification was due to the need for significant organizational adaptations and skills upgrading. Just as economic observers pushed for the cultivation of the electrical engineering discipline, the pioneers of military electrification pushed for more military electricians. In 1890, Fiske proposed the formation of a corps of naval and military electricians to assist the army and navy. It was not until eight years later during the Spanish War when Captain Eugene Griffin, VP of General Electric, formed such a corps of electricians and mechanical engineers.[113] This volunteer corps of one thousand engineers served as the foundation for the work of U.S. electricians and engineers in the First World War.[114]

---

[108] National Army Museum, n.d.
[109] MacLaren 1961, 4.
[110] MacLaren 1961, 3. This was a growth in the cost of components that mirrors the growing cost of software in modern fighter jets.
[111] Rosen 1994, 190-191.
[112] This estimate is based on U.S. government interrogations of operators of German antiaircraft guns and German radar scientists. Rosen 1994, 198.
[113] Fiske 1919, 130.
[114] Fiske 1919, 239.

Organizational barriers also prevented the fast uptake of various military applications of electricity. Take, for instance, the U.S. Navy's delayed adoption of wireless telegraphy (radio communications). Despite the opportunity to employ radio communications as early as 1899, the U.S. Navy took about fifteen years to fully integrate the radio into its operations, as senior naval officers saw the radio as a direct threat to their authority onboard ships. This ties back to how GPTs often require radical structural shifts like the redesign of factories. Before the radio, ship captains and fleet commanders were masters of their domain; but advances in electrical communications enabled military bureaucrats and land-based rivals to challenge their authority.[115] Taylor concludes, "It would take World War I, combined with several acts of Congress and executive orders by President Wilson, to finally force the shift to a radio navy."[116]

## COMPARISON TO SUBMARINES AS A NON-GPT

Not all technologies interact with military systems in the same way. Along all three features of GMTs, the evolution of submarines in military affairs differed from military electrification.[117] First, advances in underwater submersion technology affected military effectiveness through a much narrower pathway, largely in the application of submarines as weapons platforms.[118] During the late 19th century and early 20th century, underwater submersion technology had very few civilian applications, resulting in limited effects on industrial productivity.[119] Lastly, there was a relatively short delay between the introduction of the first modern submarines, which occurred around the turn of the 20th century, and their impact on military effectiveness, which was apparent in the years before World War I.[120]

---

[115] Taylor 2016, 192. See Douglas 1985.
[116] Taylor 2016, 192.
[117] We only briefly survey the impact of submarines on military affairs due to both space constraints and the fact that this is relatively well-covered terrain. Electricity, on the other hand, is relatively understudied. Thus, we focus the bulk of our empirical presentation on piecing together the historical data for the electrification of warfare. For military histories of submarines, see Coté Jr. 2003; Lautenschlager 1986. For an extended discussion of submarines as a non-GMT, see the online appendix.
[118] The only other substantial application was in reconnaissance. In World War I, the British did use submarines as important supplements to radio intelligence, as the development of new wireless transmitters allowed patrol submarines to relay information back to the larger fleet. Hezlet 1975, 133.
[119] More recent advances in submarine technology do benefit radar and ocean imaging capabilities.
[120] The *Holland*, introduced by the U.S. Navy in 1900, is generally considered the first modern submarine. Coté Jr. 2003, 5. By 1913, all of the leading navies had substantial submarine fleets. Crisher and Souva 2014.

INDUSTRIAL DEPENDENCY HYPOTHESIS

The broad applicability of GPTs across many sectors, combined with the fact that the civilian realm presents many more application scenarios than the military realm, means that the military will grow more dependent on the civilian sector as the GPT develops. While this effect exists to some degree for many (but not all) dual-use technologies,[121] it is significantly enhanced with the GPTs due to their broad applicability across many sectors.

This was the case with electricity. The amount of money invested in the research and development of electrical engineering departments and electric companies across the United States was "many times greater than that invested in all [the U.S.'s] ships put together."[122] The civilian economy attracted the brightest minds in electricity. As Fiske recounts, "The principal difficulty that electricity has had to meet in our Navy has been...that most of those who have become proficient (enlisted men as well as officers) have gone into civil life, and we find them distributed among the various colleges and electrical enterprises of the land...so far as helping the Navy goes, their services are lost."[123]

The widespread applicability of electricity across so many military domains highlighted the need for military-wide talent upskilling. "Every soldier or sailor, if he desires to make his mark, must be something of an electrician, for there seems to be no limit to the useful applications of the galvanic spark in battle," wrote Pritchard in an 1877 *Nature* article.[124] In the navy, electric applications significantly enhanced the complexity of ships, which gave greater advantage to the navy that devoted sufficient time to education and mastering the difficulties of these new systems.[125]

The U.S. Congressional Amalgamation Act of 1899 represents the most dramatic demonstration of how militaries faced pressures to adapt their skill base to GPT trajectories. The act merged the separate

---

[121] Military aircraft technology has become less dependent on civilian aircraft technology. The connections between nuclear weapons technology and civilian nuclear technology are also muted.
[122] Fiske 1896, 424.
[123] Fiske 1896, 327.
[124] Pritchard 1877, 281–282.
[125] Fiske 1896, p. 327

Engineering Corps with naval line officers, setting the principle that all naval officers needed to be engineers in some sense. One study of the effects of the Amalgamation Act on technology-skill complementarities concluded that it was "likely important for the Navy to continue to implement engineer-intensive general purpose technologies."[126]

Within the broader case of the electrification of military affairs, we exploit variation between the British and Russian militaries in their adoption of electrical military innovations to test our proposition about the leveling effect of GPTs.[127] As referenced in earlier sections, the British were very successful in building an effective system of wireless military communications and in manipulating other countries' wireless communications in World War I. Compared to its Russian competitor, Britain's military could draw from a more robust industrial base in electricity (Figure 3). This was encapsulated in the British advantage in signals intelligence in fighting "the world's first electronic war," a campaign that "used the best technology and science and electrical engineers of the day."[128]

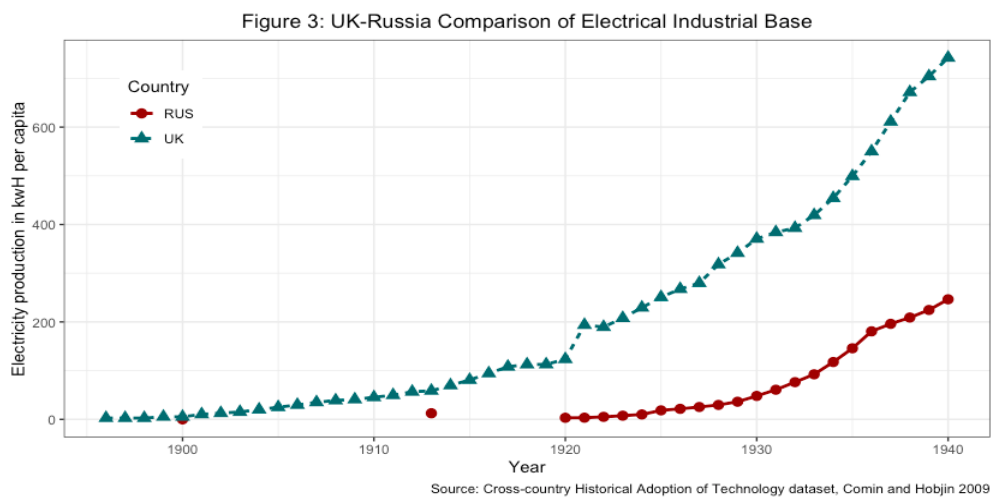

Figure 3: UK-Russia Comparison of Electrical Industrial Base

Source: Cross-country Historical Adoption of Technology dataset, Comin and Hobjin 2009

In the following sections we contrast this experience with the Russian case. Before World War I, Russia was a first rank power and perceived by others to be growing rapidly in power.[129] However, its military was a technological laggard with respect to electrical communications. In fact, according to one comparison of the signal communication systems of the British and Russian militaries at the onset of World War I, Britain and Russia were at complete opposite ends of the spectrum.[130]

This failure came to the fore at the Battle of Tannenberg between Russian and German forces at the onset of World War I, to dramatic effect. The Russian army suffered a massive defeat, and 100,000 men were captured.[131] According to George I. Back, who served as the chief signal officer of the Mediterranean theater for the U.S. Army, and George Thompson, a military historian, Russia's setback at Tannenberg "was largely due to an almost total lack of signal communication."[132] Because the Russian military lacked both the electrical equipment and the requisite technical knowledge base for using these devices and encrypting electrical communications, the Germans had access to detailed Russian communications and marching orders.[133]

One key factor behind the Russian military's relative failure with adopting electricity was its weakly developed industrial base in electricity. The lack of a unified technical profession and the lack of skilled personnel functioned as a bottleneck on Russian military electrification.[134] With no leading electrical companies of their own, the Russian electrical engineering sector depended on foreign suppliers for critical components such as high voltage transformers and measuring instruments.[135] When imports were cut off amidst war, Russian industry could not independently manufacture this equipment, even with support from the Russian war industry committee.[136]

If industrial dependency holds, then even a strong, technologically savvy military cannot bring about a GMT on its own. Before World War I, the Russian military tried to promote electrification by whatever means necessary. It sponsored research and travel, trained scientists, pushed for the creation of a domestic industry, gathered information about the latest developments in electric technology, functioned as a testbed for materials and systems, and also served as the largest procurer of electricity.[137] While the Russian military did achieve success in some limited, early applications of electricity, this success did not extend to electrification as a whole. Russian electric mine technology, for instance, was relatively advanced. Russian electric mines damaged two British ships in the Crimean War, and the Japanese sent a delegation to the Russian Mine School in 1877 to learn from Russian officers about the production, maintenance, and use of electric mines. However, this competent pool of workers was "one of the few such groups in Russia."[138] Unfortunately for the Russian military, taking full advantage of a GPT requires much more than skill competencies in a few application sectors; it requires a robust industrial base familiar with spreading the GPT.

---

[130] Back and Thompson 1998.
[131] Headrick (1991, 156) describes the Russians as "the worst off" with respect to radio equipment.
[132] Back and Thompson 1998.
[133] One of the two main Russian armies at Tannenburg only had twenty-five telephones and a handful of manual Morse Code machines. Jackson 2002, 4.
[134] Coopersmith 1992, 97-101
[135] Graham 2013, 28-29; Coopersmith 1992, 97-101
[136] Coopersmith 1992, 104. For another case of slow Russian military electrification, in which the Russian Navy was hampered by its reliance on German wireless equipment in the Russo-Japanese War. see Lee 2012, 25-28; Hezlet 1975, 42-49.

What about other factors besides industrial dependency? For instance, Russia's failure to exploit the electricity-based GMT could stem from the fact that the Russian economy as a whole was more underdeveloped than the British. The divergent rates of military electrification could be due to differences in the two militaries' organizational and cultural attitudes toward emerging technologies. To control for alternative explanations, we examine trends in submarine technologies, which do not fulfill the characteristics of a GPT. If Russia was able to keep pace with Britain adopting submarines, then it is less likely that general economic, organizational, and cultural factors were driving differences in military electrification.

In line with these expectations, we find that the Russian navy was closer to parity with the British in terms of submarine capabilities than military electrification. Before the outbreak of WWI, the Russian navy was equipped with 18 diesel-electric submarines,[139] of which the Bars and Morzh class submarines were on par with the best foreign counterparts. Russian submarine engineers were also experienced and knowledgeable, and Russian submarines conducted effective operations in the Black Sea during the war.[140] Overall, there was significantly less of a gap between Russian and British submarine capabilities than the gap between Russia's GMT in electricity and Britain's GMT in electricity.

---

[137] Coopersmith 1988.
[138] Coopersmith 1988, 298.
[139] Crisher and Souva 2014
[140] Polmar and Noot 1991, 28-29; Muraviev 2015, 86-87

Since advances in submarine capabilities did not generate a GMT, a military's connection to a robust industrial base should be less significant for explaining which militaries are differentially advantaged by submarines. Like Russia's electrical base, Russia's civilian shipbuilding industry was weak, but this weakness was not as significant in determining the military effectiveness of Russian submarine capabilities.[141] Developing effective submarine capabilities required a relatively narrow talent base skilled in operating high-speed reciprocating machinery.[142] Russia struggled more with electrification because of the demand for general upskilling drawing on a broad industrial base connected to electrical advances. In fact, one of the main shortcomings of Russian submarines was linked to the Russian military's weakness in electrification. In World War I Russia's dependence on Germany for diesel-electric engines for stronger propulsion was exposed when war broke out.[143]

While this is a promising first step in the study of GMTs, the generalizability of a single case is limited. It is also important to specify scope conditions when extending the implications of the electricity case to other possible GMTs. Whether the lessons from history about the effects of electricity on military affairs can apply to current developments in AI depends not just on whether the technical properties of AI are comparable to those of electricity but also on the congruence of fit between the current period and the nature of military competition and dynamics of the international system in the late 19th century.

---

[141] Russian shipyards were notoriously slow. Hauner 2004, 94.
[142] Nimitz 1916, 487
[143] Polmar and Noot 1991, 28-29; Muraviev 2015, 86-87. Russia quickly realized the strategic potential of submarines as new combat platforms, and Russian submarines played an effective role during the Russo-Japanese War and World War I. In World War I, the Russian submarine fleet made the German-Turkish naval forces sustain significant losses, and the Soviet submarine school, after the war, emerged as one of the world's leading centers of submarine warfare.

# IV. Conclusion and Lessons for AI

The evidence from the electricity case is compatible with GMT theory. Similar to the trajectory of electricity in the economic domain, electricity found widespread military applications only after a protracted period of gestation. In addition to directly boosting military effectiveness through a broad array of military innovations, electricity also indirectly transformed military power through stimulating industrial productivity. Institutional adaptations (both civilian and military) to widen the talent base associated with electricity, especially in electrical engineering, were crucial to help certain militaries benefit more from the GMT.

These findings should be especially relevant for discussions of how military affairs will be transformed by AI — the "next GPT."[144] Based on the features of GMTs and the industrial dependency hypothesis, we can state some predictions for how AI will affect the future of warfare. Our arguments contribute to existing discussions about the effect of AI in military affairs, which often emphasize AI as a Revolution in Military Affairs that rising military powers can take advantage of to leapfrog the U.S. in weapons capabilities.[145]

While AI and electricity are both GPTs, they differ along many other relevant characteristics. Autonomy, for instance, is a distinctive characteristic of some AI systems. Moreover, since the late 19th century, there have been changes in the nature of military competition and innovation. All of these contextual factors need to be taken into account when generalizing our GMT-related findings to the AI case. Still, our analysis can provide an initial guide for comprehending the impact of AI on military affairs, akin to how studies of electricity's effect on economic transformation have improved discussions about the impact of computers on productivity.[146]

First, speculation about how AI will transform military affairs places excessive emphasis on the narrow effects of weapon systems. Autonomous weapons systems have drawn a large share of the attention from policymakers and scholars alike, as have narratives of an "AI arms race."[147] U.S. defense intellectuals highlight how China could take advantage of AI-enabled hypersonic missile systems to leapfrog U.S. military power.[148] As Elsa Kania notes, the Chinese military's focus on "trump card" or "assassin's mace" weapons that can counter U.S. capabilities "will likely persist in the PLA's approach to AI.[149]

---

[144] Trajtenberg 2018
[145] For more on RMAs see Krepinevich 1994; Krepinevich 2002.
[146] See, for example, David 1990

In contrast, a GMT approach emphasizes the accumulation of AI-enabled improvements across many military systems. This impact pathway will likely involve significant upgrades to weapons capabilities, as was the case with electricity and centralized fire control. On the whole, though, effects of AI advances in other military domains, including communications, cyberspace operations, intelligence, information and psychological operations, logistics, strategic decision-making, etc. will be more consequential.[150] Moreover, the focus on AI weapons neglects the indirect pathway of influence through AI's potential to upgrade the productive capabilities of the overall industrial base. In particular, the intelligentization of manufacturing lines (smart manufacturing) could have significant follow-on effects for military readiness.

Second, existing conjectures about the impact of AI on military affairs severely underestimate the timeframe for when substantial effects will occur. One highly-cited article on AI and national security argues, "The amount of progress AI technology is poised to make over the next 10-20 years should lead the Department of Defense to revisit those assumptions (about spending priorities on aircraft and naval platforms)."[151] Payne also estimates, "(T)he rapid progress in AI research, especially of hybrid approaches that utilise multiple AI techniques, along with increasingly powerful hardware on which to run algorithms, suggests the potential for AI to significantly affect existing military activities in the short to medium term…".[152] This is reflective of a broader tendency in strategic thinking: to conflate rapid progress in a

---

technological field, which is characteristic of GPTs, with rapid adoption across military applications, which is uncharacteristic of GPTs.

According to the GMT theory, the most consequential impacts of AI on military effectiveness will occur only after a long period of gestation. Economists have already begun to model implementation lags in the effects of AI on economic productivity.[153] A similar extended trajectory will apply in the military realm. The current wave of AI development started with the deep learning revolution in the early 2010s, so if AI follows the same timeline as electricity, a prolonged period of gestation could extend until around the 2050s.[154] In addition, since the development of AI is still in its early stages, the foreseeability of its military applications is very limited. Even the most astute observers of military transformation at the turn of the 20th century, twenty years after the introduction of the electric dynamo, could not envision how the technology would transform military affairs. As only a decade has passed since critical breakthroughs in deep learning, any attempt to foreordain the ultimate strategic impacts of AI should be met with deep skepticism. Our imaginations — to borrow language from the ordnance engineer quoted earlier — are not sufficiently elastic.

Lastly, the GMT theory supplements and modifies existing thinking about the rate of diffusion of military applications of AI and the effect of AI on the military balance of power. Some scholars argue that if military advances in AI continue to be closely linked to civilian applications, then military AI capabilities will rapidly diffuse to other countries.[155] Other factors commonly cited include the financial and organizational requirements for adopting military AI technology. Informed by a historical perspective of GMTs, we view "military AI technology" as not a singular technological innovation but part of a GPT trajectory, which encompasses a broad distribution of technological applications. Just like the organizational requirements for adopting wireless telegraphy were different from those required to adopt searchlights, the adoption capacity for different applications of AI will vary.

---

[153] Brynjolfsson et al. 2017
[154] This expectation can be affected by other factors, including the possibility that the general process of technological adoption is accelerating. Some evidence indicates that the waiting time for a significant productivity boost from a new GPT has decreased over time. Crafts 2004.
[155] Drezner 2019; Horowitz 2018.

To more fully account for how AI advances will differentially advantage certain militaries, more attention should go to factors that apply across the broad front of a GPT trajectory.[156] We highlight the significance of a state's industrial capacity to provide AI infrastructure and skilled labor to militaries. Specifically, militaries able to draw from a wide skill base in AI will better exploit the AI-based GMT. Crucially, the talent base required for AI differs from the talent base required for other revolutionary dual-use technologies like nuclear power. GMT theory suggests that military linkages to a wide base of AI engineering talent, rather than star researchers or cutting-edge technical capabilities, are crucial to adapting generalized models to a variety of specific military applications.[157]

The three great influences on naval warfare of the 20th century were the aircraft, the submarine, and electricity. We have shown that electricity is not like the others. It powered a GMT. In parallel characterization, Horowitz identifies three key technologies that could reshape the future of warfare in the 21st century: cyber, drones, and AI.[158] As plausibly the defining GPT of our century, AI is not like the others. After all, it's the new electricity.

---

[156] It is important to note that specialized military applications of AI, with few linkages to the commercial domain, do exist and will emerge. See Horowitz 2020b.
[157] Payne 2018; Ryseff 2020. Our conclusions diverge from existing discussions in the U.S. that focus on the DoD's access to top elite talent in top tech firms.
[158] Horowitz 2020a; Montgomery (2019, 324) writes that "AI is often identified as the emerging technology that could most influence military power."